\documentclass{article}
\usepackage{arxiv}

\usepackage[T1]{fontenc}
\usepackage[utf8]{inputenc}
\usepackage{amsmath}
\usepackage{cite}
\usepackage{url}
\usepackage{graphicx}
\usepackage{color}
\usepackage{amsfonts,booktabs}

\title{Variable-Rate Harmonic-Percussive Time-Scale Modification with Real-Time Playback in Python}

\author{
Sayema Lubis$^{1}$, Clark Peng$^{2}$, Jared Carre\~no$^{1}$, TJ Tsai$^{1}$ \\
$^{1}$Harvey Mudd College, Claremont, CA, USA \\
$^{2}$University of California, Los Angeles, CA, USA \\
}

\date{}

\begin{document}

\maketitle

\begin{abstract}
Time-scale modification (TSM) has a number of open-source implementations, but these are designed almost exclusively for offline use, in which a recording is processed at a fixed rate and written out ahead of time. Applications such as automatic musical accompaniment require a different setting, which we call variable-rate playback: the recording to be stretched is known in advance, but the playback rate is not, and must change continuously in response to a live performer. The few implementations that generate output in real time are written in C++ and optimized for speed rather than for ease of modification, experimentation, and integration with the primarily Python-based research ecosystem. This paper describes a Python implementation of the widely used harmonic-percussive TSM method for the variable-rate playback setting.  Harmonic-percussive separation is performed offline as a preprocessing step on the known input recording, while synthesis and playback are carried out in real time with a time-scale factor that may change at every frame. We further propose a family of variants that reduce runtime by replacing the phase vocoder's analysis-stage FFT and instantaneous frequency calculations with lookups into precomputed tables. Subjective listening tests with 24 participants (1114 pairwise ratings) show that these approximations become perceptually indistinguishable from the exact implementation once the precomputed hop size is sufficiently small, while reducing total runtime by roughly half. We characterize the resulting tradeoffs among precomputation, runtime, memory, and perceptual quality to guide algorithm selection, and we release our implementation as an open-source package.
\end{abstract}

\section{Introduction}
\label{sec:intro}

Time-scale modification (TSM) is the process of changing the speed or duration of an audio signal without altering its pitch or introducing any unwanted artifacts.  TSM algorithms are used in all modern digital audio workstations and music production software, and are readily available in many commercial products.  In the research community, there are a number of open source implementations of TSM for research, but these packages heavily focus on TSM in offline scenarios.  This paper describes how the widely used (offline) TSM algorithm based on harmonic-percussive separation \cite{driedger2016review, driedger2013improving} can be adapted for variable-rate playback, generating output in real time at a rate that can change continuously, along with a family of approximations that reduce runtime.  Our goal is to offer a resource that enables real-time TSM applications and integrates seamlessly into the existing landscape of research tools.

Previous work on TSM falls into four main categories.  The first category are time-domain approaches based on Overlap-Add (OLA), in which windowed analysis frames from the original signal are re-combined with a different hop size in the synthesis stage.  OLA handles percussive sounds well but cannot maintain local periodic structures, resulting in phase jump artifacts.  To address this issue, previous methods allow for minor shifts in the analysis frame positions (WSOLA) \cite{verhelst1993overlap} or synthesis frame positions \cite{roucos1985high, moulines1990pitch, laroche1993autocorrelation} to maintain periodic structures.  The second category are frequency-domain approaches based on the Phase Vocoder (PV) \cite{flanagan1966phase, laroche2002improved, portnoff2003implementation}.  With a PV-based approach, an STFT is calculated on the original signal, the phase components of the STFT coefficients are modified to avoid phase jump artifacts within each frequency bin, and the modified STFT is inverted back into a time-domain signal.  Because the PV preserves periodic structures, it works well with harmonic sounds but loses vertical phase coherence, resulting in transients that sound fuzzy and less distinct \cite{laroche1997phase}.  To mitigate these artifacts, some methods restore vertical phase coherence on a regular basis by copying an analysis frame unmodified to the output and using a PV approach in between \cite{kraft2012improved, moinet2011pvsola, dorran2006hybrid}.  Other approaches mitigate the artifacts using phase locking  \cite{laroche2002improved, duxbury2002improved}, in which time-frequency bins with a magnitude peak are updated with a PV approach but neighboring frequency bins are locked to the phase of the peaks.  The third category are approaches that decompose signals into 2 or more components and process each component differently.  These approaches tend to have higher perceptual quality at the cost of higher complexity.  Some approaches detect and preserve transients in the signal, while using a PV approach for the non-transient regions or components \cite{grofit2007time, nagel2009novel, duxbury2002improved}.  Some approaches decompose signals into sine, transient, and noise components, and use appropriate TSM approaches for each component \cite{moliner2024noise, damskagg2017audio, verma1998time}.  Of particular note, Driedger and Mueller \cite{driedger2016review, driedger2013improving} decompose signals into percussive and harmonic components, process them respectively with OLA and PV methods, and recombine the outputs.  This method is one of the most widely used methods in the MIR community, and will serve as a focal point for our current study.  The fourth category are recent works that explore the use of neural synthesis for TSM \cite{chu2022audio, fierro2023extreme, jang2024diffatsm}.  Because our interest is on real-time implementations that run on general-purpose hardware, we will not focus on such methods in this paper.

It is useful to consider the landscape of software resources for TSM.  As mentioned above, there are many real-time implementations of TSM methods in commercial products, but these are not conducive to research due to their restrictive licenses.  There are several open-source implementations of various TSM methods, such as TSM Toolbox \cite{driedger2014tsm} (implemented in Matlab) and its python re-implementation libtsm \cite{rosenzweig2021adaptive}, librosa \cite{mcfee2015librosa}, PyTSMod \cite{yong2020pytsmod}, and AudioStretchy \cite{audiostretchy}.  All of these implementations are designed for offline use in python or Matlab.  To the best of our knowledge, there are only two open-source implementations (both in C++) that are real-time capable: Soundtouch \cite{soundtouch} and Rubber Band \cite{rubberband}.  Soundtouch uses a ``WSOLA-like time-stretching routine'' while Rubber Band implements a ``block-based phase vocoder with phase resets on percussive transients''.  The authors of Rubber Band mention that their method ``does not implement any single complete published method'' and that some aspects of their method have not been published anywhere.  These libraries have two main drawbacks.  First, they are both implemented in C++ for speed, which makes it harder to modify, adapt, and integrate with existing python tools.  Second, they use methods that are either not state-of-the-art (Soundtouch) or have not been vetted by the research community (Rubber Band).

Our goal in this work is to develop an open-source implementation of TSM that (a) can support dynamically changing TSM factors in real-time, (b) implements or approximates a state-of-the-art (offline) method that is widely used in the research community, and (c) is easy to adapt, modify, and integrate within the ecosystem of primarily python-based research software.  Our goal is to provide a resource that facilitates research on applications involving real-time TSM, such as an accompaniment system that plays back a pre-recorded accompaniment track that is adjusted to match a user's live playing.

This paper has three main contributions.  First, we describe the design of a real-time version of a widely used (offline) time-scale modification algorithm based on harmonic-percussive separation \cite{driedger2016review, driedger2013improving}.  This implementation allows a user to change the tempo of a recording in real-time, while preserving the benefits of a harmonic-percussive approach.  Because runtime is a critical factor in real-time systems and depends on the available hardware, a one-size-fits-all approach to algorithm design is overly restrictive.  Accordingly, we also propose several variants to reduce runtime by approximating the original algorithm.  Second, we demonstrate that it is possible to perform state-of-the-art real-time TSM natively in python/numba.  Implementing this in python makes it easy to modify, extend for research purposes, and integrate with the existing landscape of python-based research tools.\footnote{Code for this paper can be found at https://github.com/HMC-MIR/TSMRealTime.}  Third, we conduct a series of listening experiments to compare the perceptual quality of the algorithm and its variants on a range of music.  Based on these results, we characterize the tradeoff between perceptual quality, runtime, and memory requirements for these variants, thus providing guidelines for a designer to make an appropriate selection.

\section{System Description}
\label{sec:systemDescr}

In this section, we describe a real-time implementation of a TSM algorithm based on harmonic-percussive separation (HPS), which was originally designed for an offline setting \cite{driedger2013improving} (Section 2.1).  We will refer to this system as `HPS-TSM-Realtime` for convenience.  Because runtime is a critical factor in real-time systems, we also propose several variants that are designed to approximate HPS-TSM-Realtime with reduced computation (Section 2.2).  We will refer to these variants as `HPS-TSM-Approx`.  This provides flexibility to select an appropriate real-time TSM algorithm based on the hardware constraints of a particular application.

\subsection{HPS-TSM-Realtime}
\label{subsec:hps-tsm-realtime}

The HPS-TSM-Realtime system has three steps.  The first step is to separate the input waveform $x[n]$ into a harmonic component $x_h[n]$ and a percussive component $x_p[n]$.  This step is done in an offline manner as a preprocessing step, and uses the same approach as in \cite{driedger2016review}, which is based on median filtering of the magnitude spectrogram.  The second step is to process $x_h[n]$ with a phase vocoder (PV) method and $x_p[n]$ with an Overlap-Add (OLA) method.  This arrangement takes advantage of the strengths of each TSM procedure, while minimizing their artifacts.  Finally, the outputs of both TSM procedures are mixed together and sent to the speaker.  The key difference compared to \cite{driedger2016review} is that the TSM is applied in real-time using a TSM factor $\alpha_t$ that can change dynamically.  Because $x_h[n]$ and $x_p[n]$ are computed from the full input recording ahead of time, the harmonic-percussive separation is an offline preprocessing step; throughout this paper, ``real-time'' refers to the synthesis and playback stages, which generate output on the fly at a rate that can change dynamically.

\begin{figure}[t]
	\centering
	\includegraphics[width=0.72\linewidth]{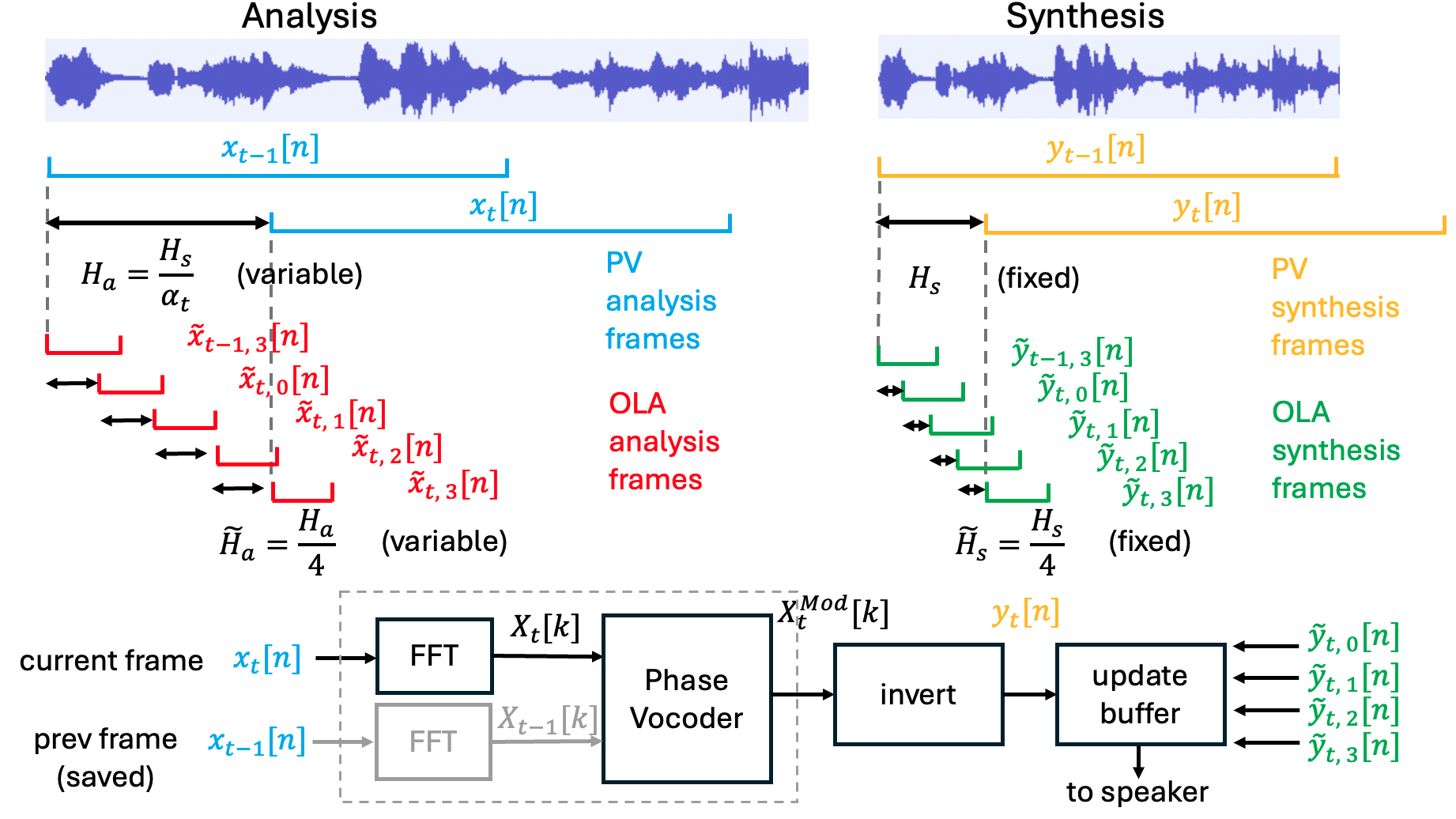}
	\caption{Overview of real-time implementation of the widely used (offline) TSM method based on harmonic-percussive separation.  We explore ways to approximate the gray dotted box with table lookups in order to reduce computation.}
	\label{fig:systemOverview}
\end{figure}

\subsection{Real-time Phase Vocoder}
\label{subsec:phaseVocoder}

Figure~\ref{fig:systemOverview} shows an overview of the real-time HPS TSM algorithm combining both PV and OLA methods.  The input signals $x_h[n]$ and $x_p[n]$ are available in their entirety, while the output is generated in real-time.  In this subsection, we give a brief overview of the PV component.  Each iteration of the PV component has 5 steps.

The first step is to determine the location of the current PV analysis frame (shown in blue as $x_t[n]$).  The hop size between the previous analysis frame $x_{t-1}[n]$ and the current analysis frame $x_t[n]$ is calculated as $H_a = \frac{H_s}{\alpha_t}$, where $\alpha_t$ is the current TSM factor and $H_s$ is the PV synthesis hop size.  $H_s$ is fixed and set to $\frac{N}{4}$, where $N$ is the length of the PV analysis \& synthesis windows.  In our experiments, we used a window size of $N=2048$ samples with a sampling rate of 22050 Hz.

The second step is to compute short-time Fourier Transform (STFT) coefficients.  The current analysis frame $x_t[n]$, $n=0,1,\dots, N-1$ is windowed with a Hann window, and the STFT coefficients $X_t[k]$, $k=0,1,\dots, N-1$ are computed as:
\begin{equation}
	X_t[k] = \sum_{n=0}^{N-1} x_t[n] \cdot w[n] \cdot e^{-j 2 \pi k n/N}
\end{equation}
For notational convenience, we can express the STFT coefficients in terms of their magnitude and phase $X_t[k] = |X_t[k]| e^{j \phi_t[k]}$, $k=0,1,\dots,N-1$.

The third step is to use the phase vocoder method to modify the phase components $\phi_t[k]$, $k=0,1,\dots, N-1$.  This is done in two steps.  First, the instantaneous frequency estimate $F^{IF}_t[k]$ is calculated by taking the nominal frequency (in rad/s) of each bin $F_{nom}[k] = 2 \pi k F_s/N$, and then applying a correction term to be consistent with the observed phases in the current ($\phi_t[k]$) and previous ($\phi_{t-1}[k]$) frames.  The instantaneous frequency terms are thus calculated as
\begin{equation}
	F^{IF}_t[k] = F_{nom}[k] + \frac{\Psi(\phi_t[k] - (\phi_{t-1}[k] + F_{nom}[k] \cdot \Delta_T))}{\Delta_T}
	\label{eq:instFreq}
\end{equation}
where $\Delta_T = \frac{H_a}{F_s}$ is the analysis hop size in seconds and where $\Psi$ is the principal argument function that maps a given phase to the interval [$-\pi$, $\pi$].  Note that the numerator of the rightmost term in equation~\ref{eq:instFreq} is the phase error when assuming that the instantaneous frequency is the nominal frequency $F_{nom}[k]$.  Once $F^{IF}_t[k]$ is calculated, the modified phase is calculated as $\phi^{Mod}_t[k] = \phi^{Mod}_{t-1}[k] + F^{IF}_t[k] \frac{H_s}{F_s}$, where $H_s$ is the PV synthesis hop size and $F_s$ is the sampling rate.  This formula ensures phase continuity within each frequency bin.  The modified STFT coefficients are given by $X^{Mod}_t[k] = |X_t[k]| e^{j \phi^{Mod}_t[k]}$, $k=0,1,\dots, N-1$.

The fourth step is to calculate the synthesis frame $y_t[n]$.  Following \cite{driedger2016review}, this is done by calculating the inverse FFT $x^{Mod}_t[n]$ of $X^{Mod}_t[k]$, and then calculating the synthesis frame as $y_t[n] = \frac{w[n] x^{Mod}_t[n]}{\sum_{r \in \mathbb{Z}} w[n-r H_s]^2}$, where $w[n]$ is a Hann window.  This process of inverting the STFT coefficients minimizes a squared error metric.

The fifth step is to reconstruct the output signal by adding overlapping synthesis frames into the buffer.  Because this process is interlinked with the OLA subsystem, we discuss this separately in section~\ref{subsec:updatingBuffer}.

\subsection{Real-time OLA}
\label{subsec:overlapAdd}

Figure~\ref{fig:systemOverview} also shows the overlap-add (OLA) component.  The input signal $x_p[n]$ is available in its entirety, while the output is generated in real-time.  Note that the PV and OLA window sizes are different: the window size for PV must be fairly large in order to ensure that the STFT has adequate frequency resolution, whereas the window size of OLA should be small in order to avoid transient doubling artifacts.  In our experiments, the OLA window size is $L=256$.

Each iteration of the real-time implementation of OLA has 3 steps.  The first step is to determine the location of the current OLA analysis frames.  Because the OLA analysis/synthesis frames are much smaller than the PV frames, multiple OLA frames are processed for each PV frame.  The OLA method uses a fixed synthesis hop size of $\tilde{H}_s = \frac{L}{2} = 128$ samples, which is $\frac{128}{512} = \frac{1}{4}$ of the synthesis hop size for PV ($H_s$).  Accordingly, we process 4 OLA frames for each PV frame.  To keep the OLA and PV frames synchronized, we only allow the current TSM factor $\alpha_t$ to change for each PV frame, but maintain the same $\alpha_t$ TSM factor for all 4 associated OLA frames.  Thus, the OLA analysis hop size is calculated as $\tilde{H}_a = \frac{\tilde{H}_s}{\alpha_t}$ and fixed across all 4 current OLA frames.  In Figure~\ref{fig:systemOverview}, the OLA analysis frames are shown in red as $\tilde{x}_{t,0}[n]$, $\tilde{x}_{t,1}[n]$, $\tilde{x}_{t,2}[n]$, $\tilde{x}_{t,3}[n]$, all of which correspond to the current iteration $t$ of the algorithm.  The second step is to calculate the OLA synthesis frames.  Each synthesis frame is calculated as $\tilde{y}_{t,i}[n] = \tilde{x}_{t,i}[n] \cdot w[n]$, $i=0,1,\dots,3$, where $w[n]$ is a Hann window.  The third step is to reconstruct the output signal by adding the overlapping OLA synthesis frames.  Because we are computing the output in real-time, this is accomplished by adding all 4 current OLA synthesis frames into the audio buffer.  This process is described in the next subsection.

\subsection{Updating Buffer}
\label{subsec:updatingBuffer}

\begin{figure}[t]
	\centering
	\includegraphics[width=0.72\linewidth]{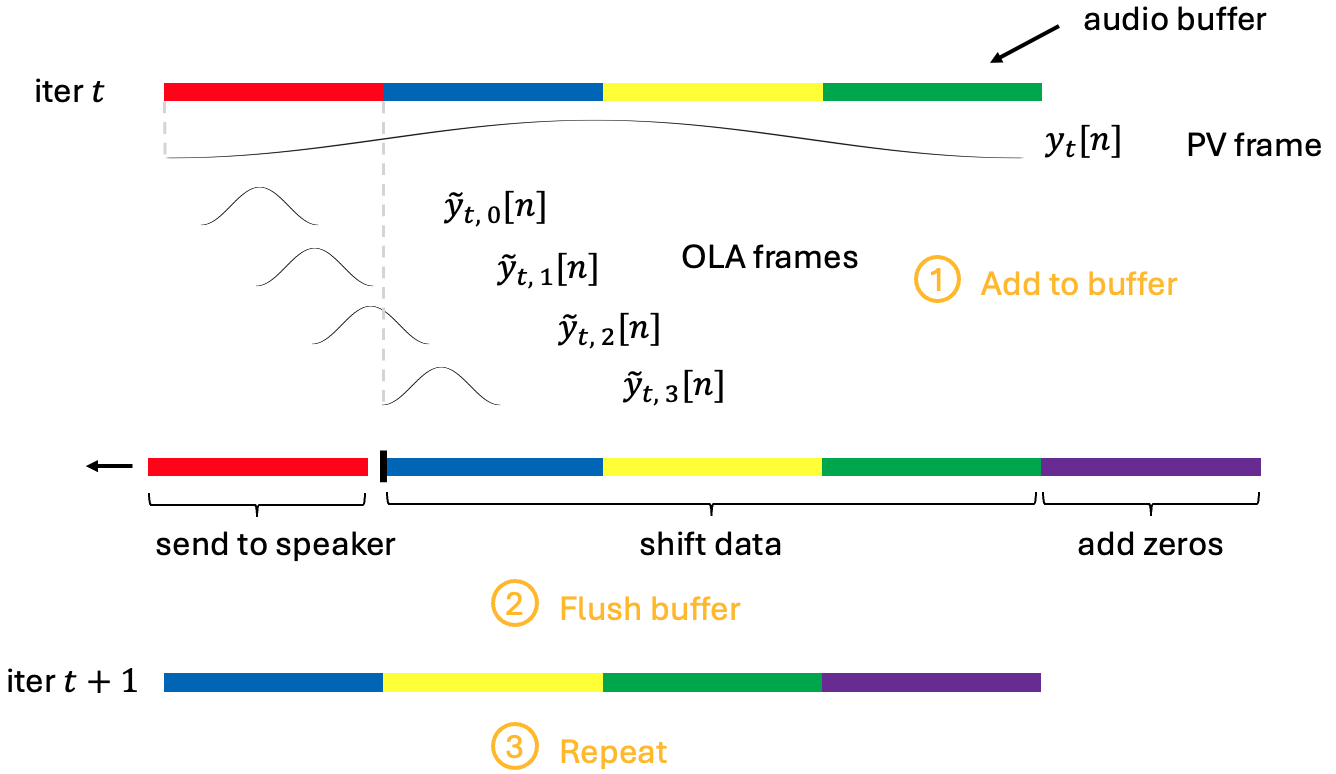}
	\caption{Updating the audio buffer by adding in both PV and OLA synthesis frames.}
	\label{fig:updatingBuffer}
\end{figure}

Figure~\ref{fig:updatingBuffer} shows the process of updating the buffer.  The audio buffer is the length of the PV window size ($N=2048$ in our experiments) and is updated in each iteration of the algorithm.  The first step is to add the PV and OLA frames into the buffer.  The PV synthesis frame $y_t[n]$ is the same size as the buffer and therefore updates every element of the buffer.  The OLA frames $\tilde{y}_{t,i}[n]$ are much shorter ($L=256$), so each OLA frame will modify $\frac{256}{2048} = \frac{1}{8}$ of the buffer elements.  All 4 OLA synthesis frames $\tilde{y}_{t,0}[n]$, \dots, $\tilde{y}_{t,3}[n]$ are added into the buffer with a synthesis hop size of $\tilde{H}_s = \frac{256}{2} = 128$ samples.  The second step is to flush the buffer by sending the first 512 samples (i.e., the portion that will no longer be changed) to the speaker for playback, shifting the remaining buffer data to the front of the buffer, and adding zeros at the end of the buffer.  This procedure is repeated at each iteration of the algorithm.

\subsection{HPS-TSM-Approx}
\label{subsec:tsmApprox}

In this subsection, we describe a family of real-time algorithms that approximate HPS-TSM-Realtime while reducing runtime.  In conducting runtime analyses of HPS-TSM-Realtime, we found that $93\%$ of the total runtime comes from the PV component (see  Figure~\ref{fig:runtimes}).  We identified two parts of the PV pipeline that could be approximated with a table lookup.  The approximation happens in two stages: an offline preprocessing stage and an online stage.

During the offline preprocessing stage, we compute the STFT of $x_h[n]$ using a (fixed) hop size of $H_{a,pre}$ samples.  We denote the precomputed STFT as $X_{h,pre} \in \mathbb{C}^{M \times N}$, where $M$ specifies the number of frames.  Based on $X_{h,pre}$, we compute the instantaneous frequency estimates $F^{IF}_{pre} \in \mathbb{R}^{M-1 \times N}$, and then store $|X_{h,pre}|$ and $F^{IF}_{pre}$ in memory as lookup tables.\footnote{Note that, because the instantaneous frequency estimates require two adjacent frames, there is one fewer frame in $F^{IF}_{pre}$ than $X_{h,pre}$.}

During the online stage, we determine the location of the current analysis frame $x_t[n]$, as described in Section~\ref{subsec:phaseVocoder}.  Let us define the starting location of the current analysis frame as $n_{cur}$, so that $x_t[n] = x_h[n + n_{cur}]$, $n = 0, 1, \dots, N-1$.  Instead of calculating the FFT, we determine the nearest neighbor frame index $m_{nn} = \text{round}(\frac{n_{cur}}{H_{a,pre}})$ and simply perform a table lookup $|X_{h,pre}[m_{nn},:]|$.  Similarly, instead of calculating the instantaneous frequency estimates, we determine a lower bounding frame index $m_{lb}=\text{floor}(\frac{n_{cur}}{H_{a,pre}})$ and perform a table lookup $F^{IF}_{pre}[m_{lb},:]$.  We then modify the STFT phase component and invert as usual.  In short, we approximate the FFT and instantaneous frequency calculations with their nearest neighbors in $|X_{h,pre}|$ and $F^{IF}_{pre}$.

The flexibility in the algorithm comes from the choice of the precomputed STFT analysis hop size $H_{a,pre}$.  We will characterize the effect of $H_{a,pre}$ on computation, memory, and perceptual quality in Section~\ref{sec:results}.  Note that when $H_{a,pre} = 1$, the approximation is perfect but requires an exorbitant amount of memory/computation.

\section{Experimental Setup}
\label{sec:expSetup}

In this section, we describe a series of listening tests that measure the perceptual quality of HPS-TSM-Realtime and its approximations.

We evaluated six different systems.  The first system is the baseline HPS-TSM-Realtime algorithm (denoted as `HPS Baseline' in our results).  This implements the offline algorithm described in \cite{driedger2016review} without any approximations.  The next five systems are approximations of HPS-TSM-Realtime with different settings of $H_{a,pre} = 2048, 1024, 512, 256, 128$.  For convenience, we consider these values as a fraction of the FFT size $N=2048$, so that the normalized hop sizes are $1.0$, $0.5$, $0.25$, $0.125$, and $0.0625$ times the analysis window length.  These systems are referred to in our results as `HPS-Approx $\beta=1.0$', `HPS-Approx $\beta=0.5$', etc.

Our subjective listening tests are pairwise comparisons with the baseline HPS-TSM-Realtime system.  We constructed a simple graphical user interface (GUI) that presents the listener with two audio widgets.  One of the audio widgets plays a short segment of music that is time-scale modified in real-time with the baseline HPS-TSM-Realtime algorithm.  The other audio widget plays the exact same segment of music that is time-scale modified in real-time with one of the 5 systems.
Both audio widgets use the same fixed TSM factor throughout the duration of the recording.  For each trial (i.e. pairwise comparison), a fixed TSM factor is selected between [0.5, 2] where sampling is done uniformly on a log scale.  The time-scale modification thus ranges between speeding up and slowing down by a factor of 2.  The listener is prompted to listen to the output of both audio widgets, and then select which one has higher quality.  Since many of the systems closely approximate the baseline HPS-TSM-Realtime system, we also allow the user to select a third option indicating that both widgets sound the same.

\begin{table}[t]
	\centering
	\small
	\caption{Results of our subjective listening tests.  Listeners were presented with a series of pairwise comparisons between each experimental real-time TSM system (shown at left) and the baseline HPS-TSM-Realtime system.  Columns 2-5 shows the total number of pairwise comparisons and the breakdown of votes among the three options (System, Baseline, or ``Sounds the same").  Column 6 shows the overall win rate of the experimental system compared to the baseline, and column 7 shows the $p$ value for a 1-sample t-test.  Results that are significant at a .05 significance level are shown with an asterisk.}
	\label{tab:results}
	\begin{tabular}{lcccccc}
		\toprule
		\textbf{System} & \textbf{Total Votes} & \textbf{Votes for system} & \textbf{Votes for Same} & \textbf{Votes for baseline} & \textbf{Win rate} & \textbf{$p$ value}\\
		\midrule
		HPS Approx $\beta=1.0$ & 157 & 10 & 17 & 130 & 0.118 & 0.000* \\
		HPS Approx $\beta = 0.5$ & 164 & 44 & 60 & 60 & 0.451 & 0.057 \\
		HPS Approx $\beta = 0.25$ & 135 & 44 & 54 & 37 & 0.526 & 0.782 \\
		HPS Approx $\beta=0.125$ & 160 & 53 & 57 & 50 & 0.509 & 0.616 \\
		HPS Approx $\beta=0.0625$ & 168 & 55 & 66 & 47 & 0.524 & 0.786 \\
		HPS baseline & 169 & n/a & 77 & 92 & n/a & n/a \\
		\bottomrule
	\end{tabular}
\end{table}

Our design of the GUI was the result of several iterations of preliminary listening tests.  In earlier versions of our GUI, we allowed the user to change a slider to adjust the TSM factor in real-time in each audio widget.  However, we observed that most participants stopped using the interactive slider after the first few minutes, and simply entered their ratings without interacting with the slider.  This had two undesirable effects: their ratings were mostly for a single fixed TSM factor rather than for a range of TSM factors (and were therefore a misleading indicator), and the default fixed TSM factor was $\alpha=1$ which does not actually change the tempo.  To counteract these undesirable effects, we chose to use a fixed but randomly selected TSM factor, as described above.  Note that the audio widgets in our GUI do not allow for dynamically changing TSM factors but are still computed in real-time.

The source audio material comes from the GTZAN dataset \cite{tzanetakis2002musical}.  GTZAN contains 10 music genres  each containing 100 30-sec audio recordings.  We select the audio source material for each trial by randomly selecting one of the 1000 recordings, and then using the first 15 seconds.

Real-time systems can be affected by many factors, such as the hardware used for computation, quality and speed of the wifi connection, etc.  Additionally, perceptual ratings of quality could be affected by the audio speaker quality or environment that the user is in (e.g., listening on headphones vs. playing through laptop speakers).  To control for many of these factors, we opted for an in-person listening test.  We recruited 24 college students to participate in the listening tests, generating a total of 1114 subjective ratings.  Participants performed the listening tests in the same physical environment, using the same computing hardware, and using the same headphones.

\section{Results}
\label{sec:results}

Table 1 shows the results of our subjective listening tests.  The leftmost column shows 6 different systems: the baseline HPS-TSM-Realtime system (`HPS Baseline') and the approximations of HPS-TSM-Realtime (`HPS Approx') with 5 different settings of $\beta = 1, 0.5, 0.25, 0.125, 0.0625$.  The next four columns show the total number of pairwise comparisons, as well as the number of votes for each option (``System", ``Sounds the Same", or ``Baseline").  The sixth column shows the win rate expressed as a percentage, calculated as
\begin{equation}
	\text{win rate} = \frac{N_A + 0.5 \cdot N_{same}}{N_A + N_{same} + N_{baseline}}
\end{equation}
where $N_A$ is the number of votes for system $A$, $N_{baseline}$ is the number of votes for the HPS baseline system, and $N_{same}$ is the number of votes for ``Sounds the same".  Note that votes for ``Sounds the same" are split between the two options.  So, for example, if we have 100 pairwise comparisons between system $A$ and the baseline with $N_A = 30$, $N_{baseline} = 50$, and $N_{same} = 20$, then the winrate for $A$ is $(30+0.5 \cdot 20)/100 = 40\%$.  Note that a win rate of 50\% indicates that the system is equally preferred to the baseline and therefore a very good approximation.  A win rate of 0\% indicates that the system is universally considered inferior to the baseline and therefore a very poor approximation.

We assess significance with a 1-sample t-test.  When considering binary comparisons between system A and the baseline, one can think of a single vote for system A as drawing a 1, a vote for the baseline system as drawing a -1, and a vote for ``Sounds the same" as drawing a 0.  In this way, we can consider votes as drawing samples from an unknown probability mass function with three possible values.  To determine if a sample mean is significantly less from 0 (i.e., whether the system is noticeably worse than our baseline), we calculate the t-statistic $t = \frac{\bar{X}-0}{\sigma/\sqrt{n}}$, where $\bar{X}$ is the sample mean, $\sigma$ is the sample standard deviation, and $n$ is the number of samples.  The seventh column indicates the $p$ value for the 1-sample t-test.  A small $p$ value indicates that the system is significantly worse than the baseline that it attempts to approximate.

There are a few things to notice about Table 1.  With $\beta = 1.0$, HPS-Approx has a very low win rate (11.8\%), indicating that the perceptual quality is much worse than the HPS Baseline.  This is a very poor approximation of the HPS TSM algorithm.  As we decrease $\beta$, we quickly approach a win rate close to 50\%, which would indicate that the system is indistinguishable from the HPS baseline.  For $\beta = 0.5$, we observe a winrate of $45.1\%$ and a $p$ value of $0.057$, which is not significant at a $0.05$ significance level.  For $\beta = 0.25$ and below, we observe win rates that are slightly more than 50\%, suggesting that we have reached the level of statistical noise.  These results suggest that $\beta = 0.25$ is a good choice for most scenarios, and that $\beta=0.5$ is a good choice if a slight perceptual degradation is acceptable.

\begin{figure}[t]
	\centering
	\includegraphics[width=0.72\linewidth]{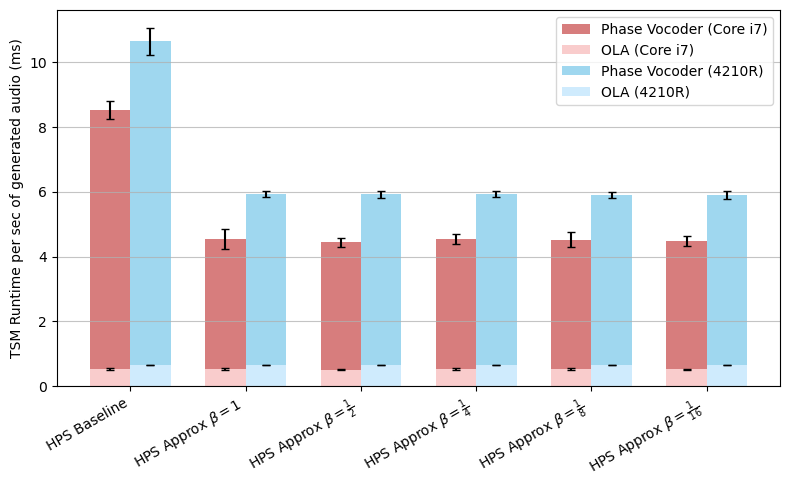}
	\caption{Runtime of various real-time TSM algorithms.  Runtimes are separated into PV and OLA components, and are reported as TSM runtime (in milliseconds) per second of generated audio.  Vertical black bars show one standard deviation above and below the mean.}
	\label{fig:runtimes}
\end{figure}

\begin{table}[t]
	\centering
	\caption{Characterizing the tradeoffs between pre-computation, runtime, memory, and perceptual quality.}
	\label{tab:tradeoffs}
	\begin{tabular}{lcccc}
		\toprule
		\textbf{System} & \textbf{Pre-computation} & \textbf{Runtime} & \textbf{Memory} & \textbf{Perceptual Quality} \\
		& \textbf{(FFTs/sec)} & \textbf{(ms/sec)} & \textbf{(MB/sec)} & \textbf{(win rate vs baseline)} \\
		\midrule
		HPS-TSM-Realtime (baseline) & 0 & 8.15 & 0 & n/a \\
		HPS Approx $\beta=1.0$ & 10.8 & 4.02 & 0.168 & 11.8\%\\
		HPS Approx $\beta = 0.5$ & 21.5 & 3.93 & 0.337 & 45.1\% \\
		HPS Approx $\beta = 0.25$ & 43 & 4.03 & 0.674 & 52.6\% \\
		HPS Approx $\beta=0.125$ & 86 & 4.01 & 1.35 & 50.9\% \\
		HPS Approx $\beta=0.0625$ & 172 & 3.97 & 2.69 & 52.4\%\\
		\bottomrule
	\end{tabular}
\end{table}

Figure~\ref{fig:runtimes} shows the empirical runtimes of the 6 real-time TSM algorithms.  We selected 50 recordings from the GTZAN dataset, applied a fixed but randomly selected TSM factor between [0.5, 2] (sampled uniformly on a log scale, as before) to each recording, and measured the runtime required to process the recording.  Since the amount of processing depends on the value of alpha (e.g., speeding up a recording takes less total computation than slowing it down), we normalize the measured runtime by the duration of the audio that is generated.  In other words, we are measuring the TSM runtime per second of audio generated.  This statistic is calculated for each audio recording, and Figure 5 shows the mean and standard deviation across the 50 recordings.  For further insight, we separately measure the runtime of the PV and OLA components.  Figure 5 shows the runtimes of all 6 systems and their breakdown into PV and OLA components, along with vertical black bars on each component to show one standard deviation above and below the means.  We ran profiling experiments with 22050Hz mono data under two different hardware configurations: one on a server with an Intel Xeon 2.40GHz processor, and one on a 2019 MacBook Pro laptop with a 2.6GHz Intel Core i7 processor.

There are three things to notice about Figure~\ref{fig:runtimes}.  First, the phase vocoder dominates the runtime of the HPS-TSM-Realtime system, accounting for about 93\% of the total runtime.  This motivated our decision to focus on reducing runtime in the PV-TSM pipeline.  Second, the HPS-Approx systems reduce the runtime of the phase vocoder component by about $50\%$, resulting in a total reduction in runtime of $45-50\%$.  Intuitively, this is what we expect since we no longer need to compute FFTs in the analysis stage but still need to compute iFFTs in the synthesis stage.  With these runtimes, we are able to perform TSM at 150 times faster than real time on consumer-grade hardware.  Keep in mind, however, that these numbers assume 22050 Hz mono data, so the runtimes would be slower for stereo data or data with a higher sampling rate.  In addition, these profiling experiments are within a regime where large memory access is not an issue.  These results are not meant to be a comprehensive evaluation of runtime, but rather a proof of concept that high-quality, real-time TSM can be performed natively in python, opening up the possibility of rapid iteration and development.  Third, the HPS-Approx systems all have roughly the same runtimes.  Thus, the real tradeoff with different values of beta is between the perceptual quality and the amount of data that needs to be kept in memory for the lookup tables.  We emphasize that these runtime reductions should not be conflated with reductions in latency.  In a streaming implementation, latency is governed by the block size and any algorithmic look-ahead, not by the per-block computation time.  As long as the system runs faster than real time, reducing runtime does not shorten latency; instead, it increases the real-time factor, providing headroom for higher sampling rates, stereo processing, or less powerful hardware.

Table 2 summarizes the main tradeoffs between pre-computation, runtime, memory, and perceptual quality.  The first column shows the 6 real-time TSM algorithms.  The second column shows the amount of computation that is required during the pre-processing stage, indicated as the number of FFTs that must be computed for each second of generated audio.  The third column shows the real-time runtime, expressed as TSM runtime (in ms) per second of generated audio.  The fourth column shows the amount of memory required to keep the lookup tables in memory, expressed as MB of memory for each second of audio in the original recording.  The memory estimates assume that each floating point number is represented by 64 bits.  The fifth column shows the win rate of the algorithm in a pairwise comparison with the baseline HPS-TSM-Realtime algorithm, which is an indicator of the perceptual quality of the algorithm.  Here, a small value indicates poor quality while a value around $0.5$ indicates that the algorithm is equally preferred.  Table 2 offers a handy reference for selecting a suitable TSM algorithm based on the costs, requirements, and constraints of the application.

There are a few things to notice about Table 2.  The HPS-Approx method with small values of beta require large amounts of pre-computation and memory.  The runtime across all values of beta is roughly the same.  The perceptual quality is at the edge of perceptibility for $\beta=0.5$ and reaches a noise floor (at least within the limitations of our listening tests) for $\beta=0.25$ and below.  Therefore, we recommend the following simple heuristic: If the runtime of HPS-TSM-Realtime is adequate, it should be used.  For applications whose hardware cannot sustain the baseline runtime, or that must reserve computation for other tasks, one can reduce the runtime by roughly half by using HPS-Approx with $\beta=0.5$ or $\beta=0.25$.  $\beta=0.5$ can be used if a slight degradation in perceptual quality is acceptable.  If perceptual quality is very important, $\beta=0.25$ can be used but requires more memory.

\section{Conclusion}
\label{sec:conclusion}

We have described how a widely used (offline) time-scale modification algorithm based on harmonic-percussive separation can be adapted for variable-rate playback, in which a known recording is played back in real time at a rate that can change continuously.  The harmonic-percussive separation is performed offline as a preprocessing step, while synthesis and playback are carried out on the fly.  We further proposed a family of approximations that reduce runtime by replacing the phase vocoder's analysis-stage FFT and instantaneous frequency calculations with lookups into precomputed tables.  We conduct a series of subjective listening tests to characterize the perceptual quality of the approximations, and we offer an analysis of the tradeoffs among runtime, memory, and perceptual quality.  Our goal is to provide a resource that facilitates research on applications involving real-time TSM, enabling more flexible and creative uses of existing offline methods.

\section*{Acknowledgments}
This material is based upon work supported by the National Science Foundation under Grant No.~2144050.  The authors thank the Claremont Graduate University Institutional Review Board for their review of this project (IRB \#5151), which was determined to not constitute human subjects research.

\bibliographystyle{IEEEtran}
\bibliography{realtimeTSM}

\begin{thebibliography}{10}
\providecommand{\url}[1]{#1}
\csname url@samestyle\endcsname
\providecommand{\newblock}{\relax}
\providecommand{\bibinfo}[2]{#2}
\providecommand{\BIBentrySTDinterwordspacing}{\spaceskip=0pt\relax}
\providecommand{\BIBentryALTinterwordstretchfactor}{4}
\providecommand{\BIBentryALTinterwordspacing}{\spaceskip=\fontdimen2\font plus
\BIBentryALTinterwordstretchfactor\fontdimen3\font minus
  \fontdimen4\font\relax}
\providecommand{\BIBforeignlanguage}[2]{{%
\expandafter\ifx\csname l@#1\endcsname\relax
\typeout{** WARNING: IEEEtran.bst: No hyphenation pattern has been}%
\typeout{** loaded for the language `#1'. Using the pattern for}%
\typeout{** the default language instead.}%
\else
\language=\csname l@#1\endcsname
\fi
#2}}
\providecommand{\BIBdecl}{\relax}
\BIBdecl

\bibitem{driedger2016review}
J.~Driedger and M.~M{\"u}ller, ``A review of time-scale modification of music
  signals,'' \emph{Applied Sciences}, vol.~6, no.~2, p.~57, 2016.

\bibitem{driedger2013improving}
J.~Driedger, M.~M{\"u}ller, and S.~Ewert, ``Improving time-scale modification
  of music signals using harmonic-percussive separation,'' \emph{IEEE Signal
  Processing Letters}, vol.~21, no.~1, pp. 105--109, 2013.

\bibitem{verhelst1993overlap}
W.~Verhelst and M.~Roelands, ``An overlap-add technique based on waveform
  similarity ({WSOLA}) for high quality time-scale modification of speech,'' in
  \emph{IEEE International Conference on Acoustics, Speech, and Signal
  Processing ({ICASSP})}, vol.~2, 1993, pp. 554--557.

\bibitem{roucos1985high}
S.~Roucos and A.~Wilgus, ``High quality time-scale modification for speech,''
  in \emph{Proc. of the IEEE International Conference on Acoustics, Speech, and
  Signal Processing ({ICASSP})}, vol.~10, 1985, pp. 493--496.

\bibitem{moulines1990pitch}
E.~Moulines and F.~Charpentier, ``Pitch-synchronous waveform processing
  techniques for text-to-speech synthesis using diphones,'' \emph{Speech
  Communication}, vol.~9, no. 5-6, pp. 453--467, 1990.

\bibitem{laroche1993autocorrelation}
J.~Laroche, ``Autocorrelation method for high-quality time/pitch-scaling,'' in
  \emph{Proc. of IEEE Workshop on Applications of Signal Processing to Audio
  and Acoustics ({WASPAA})}, 1993, pp. 131--134.

\bibitem{flanagan1966phase}
J.~L. Flanagan and R.~M. Golden, ``Phase vocoder,'' \emph{Bell System Technical
  Journal}, vol.~45, no.~9, pp. 1493--1509, 1966.

\bibitem{laroche2002improved}
J.~Laroche and M.~Dolson, ``Improved phase vocoder time-scale modification of
  audio,'' \emph{IEEE Transactions on Speech and Audio Processing}, vol.~7,
  no.~3, pp. 323--332, 2002.

\bibitem{portnoff2003implementation}
M.~Portnoff, ``Implementation of the digital phase vocoder using the fast
  fourier transform,'' \emph{IEEE Transactions on Acoustics, Speech, and Signal
  Processing}, vol.~24, no.~3, pp. 243--248, 2003.

\bibitem{laroche1997phase}
J.~Laroche and M.~Dolson, ``Phase-vocoder: About this phasiness business,'' in
  \emph{Proc. of IEEE Workshop on Applications of Signal Processing to Audio
  and Acoustics ({WASPAA})}, 1997.

\bibitem{kraft2012improved}
S.~Kraft, M.~Holters, A.~von~dem Knesebeck, and U.~Z{\"o}lzer, ``Improved
  {PVSOLA} time-stretching and pitch-shifting for polyphonic audio,'' in
  \emph{Proc. of the International Conference on Digital Audio Effects (DAFx)},
  2012, pp. 17--21.

\bibitem{moinet2011pvsola}
A.~Moinet and T.~Dutoit, ``{PVSOLA}: A phase vocoder with synchronized
  overlap-add,'' in \emph{Proc. of the International Conference on Digital
  Audio Effects (DAFx)}, 2011, pp. 19--23.

\bibitem{dorran2006hybrid}
D.~Dorran, R.~Lawlor, and E.~Coyle, ``A hybrid time--frequency domain approach
  to audio time-scale modification,'' \emph{Journal of the Audio Engineering
  Society}, vol.~54, no. 1/2, pp. 21--31, 2006.

\bibitem{duxbury2002improved}
C.~Duxbury, M.~Davies, and M.~B. Sandler, ``Improved time-scaling of musical
  audio using phase locking at transients,'' in \emph{Audio Engineering Society
  Convention}, 2002.

\bibitem{grofit2007time}
S.~Grofit and Y.~Lavner, ``Time-scale modification of audio signals using
  enhanced {WSOLA} with management of transients,'' \emph{IEEE Transactions on
  Audio, Speech, and Language Processing}, vol.~16, no.~1, pp. 106--115, 2007.

\bibitem{nagel2009novel}
F.~Nagel and A.~Walther, ``A novel transient handling scheme for time
  stretching algorithms,'' in \emph{Audio Engineering Society Convention},
  2009.

\bibitem{moliner2024noise}
E.~Moliner, L.~Fierro, A.~Wright, M.~S. H{\"a}m{\"a}l{\"a}inen, and
  V.~V{\"a}lim{\"a}ki, ``Noise morphing for audio time stretching,'' \emph{IEEE
  Signal Processing Letters}, vol.~31, pp. 1144--1148, 2024.

\bibitem{damskagg2017audio}
E.-P. Damsk{\"a}gg and V.~V{\"a}lim{\"a}ki, ``Audio time stretching using fuzzy
  classification of spectral bins,'' \emph{Applied Sciences}, vol.~7, no.~12,
  p. 1293, 2017.

\bibitem{verma1998time}
T.~S. Verma and T.~H. Meng, ``Time scale modification using a sines +
  transients + noise signal model,'' in \emph{Proc. of the Digital Audio
  Effects Workshop ({DAFX})}, 1998, pp. 49--52.

\bibitem{chu2022audio}
E.~Chu, J.-T. Chen, and C.-P. Chen, ``Audio time-scale modification with
  temporal compressing networks,'' \emph{arXiv preprint arXiv:2210.17152},
  2022.

\bibitem{fierro2023extreme}
L.~Fierro, A.~Wright, V.~V{\"a}lim{\"a}ki, and M.~H{\"a}m{\"a}l{\"a}inen,
  ``Extreme audio time stretching using neural synthesis,'' in \emph{Proc. of
  the IEEE International Conference on Acoustics, Speech and Signal Processing
  ({ICASSP})}, 2023, pp. 1--5.

\bibitem{jang2024diffatsm}
S.~Jang, Y.-J. Kim, and J.-H. Chang, ``Diffatsm: High quality adaptive
  time-scale modification using diffusion-based post-processing,'' \emph{SSRN
  preprint}, 2024.

\bibitem{driedger2014tsm}
J.~Driedger and M.~M{\"u}ller, ``{TSM Toolbox}: {MATLAB} implementations of
  time-scale modification algorithms,'' in \emph{Proc. of the International
  Conference on Digital Audio Effects (DAFx)}, 2014, pp. 249--256.

\bibitem{rosenzweig2021adaptive}
S.~Rosenzweig, S.~Schw{\"a}r, J.~Driedger, and M.~M{\"u}ller, ``Adaptive
  pitch-shifting with applications to intonation adjustment in a cappella
  recordings,'' in \emph{International Conference on Digital Audio Effects
  (DAFx)}, 2021, pp. 121--128.

\bibitem{mcfee2015librosa}
B.~McFee, C.~Raffel, D.~Liang, D.~P. Ellis, M.~McVicar, E.~Battenberg, and
  O.~Nieto, ``librosa: Audio and music signal analysis in python,''
  \emph{SciPy}, vol. 2015, pp. 18--24, 2015.

\bibitem{yong2020pytsmod}
S.~Yong, S.~Choi, and J.~Nam, ``{PyTSMod}: A python implementation of
  time-scale modification algorithms,'' in \emph{Late-Breaking Demo Session of
  the International Society for Music Information Retrieval Conference (ISMIR
  LBD)}, 2020.

\bibitem{audiostretchy}
A.~Twardoch, ``Audiostretchy library,''
  \url{https://github.com/twardoch/audiostretchy}, accessed: 2026-04-15.

\bibitem{soundtouch}
O.~Parviainen, ``Soundtouch audio processing library,''
  \url{https://www.surina.net/soundtouch/}, accessed: 2026-04-15.

\bibitem{rubberband}
B.~Quay, ``Rubber band library,'' \url{https://breakfastquay.com/rubberband/},
  accessed: 2026-04-15.

\bibitem{tzanetakis2002musical}
G.~Tzanetakis and P.~Cook, ``Musical genre classification of audio signals,''
  \emph{IEEE Transactions on Speech and Audio Processing}, vol.~10, no.~5, pp.
  293--302, 2002.

\end{thebibliography}

\end{document}